\documentclass[oneside,10pt]{amsart}
\pdfoutput=1

\usepackage{geometry}
\usepackage[T1]{fontenc}
\usepackage[utf8]{inputenc}
\usepackage{lmodern}
\usepackage[dvipsnames]{xcolor}
\usepackage{tikz,tikz-3dplot}
\usetikzlibrary{positioning,decorations.pathreplacing}

\newcommand{\sO}{\mathcal{O}}

\theoremstyle{plain}

\title{$\phi^3$ theory at six loops}
\author{Oliver Schnetz}
\address{Oliver Schnetz\\
II. Institut f\"ur Theoretische Physik\\
Luruper Chaussee 149\\
22761 Hamburg, Germany}
\email{schnetz@mi.uni-erlangen.de}

\begin{document}
\begin{abstract}
We present the renormalization functions of dimensionally regularized $\phi^3$ theory in six dimensions up to loop order six in the minimal subtraction scheme.
\end{abstract}
\maketitle

\section{Introduction}
The calculation of the renormalization functions $\beta$, $\gamma$, and $\gamma_m$ for a given Quantum Field Theory (QFT) is a classical problem in particle physics \cite{IZ}.
The beta function determines the running of the coupling while $\gamma$ and $\gamma_m$ are the anomalous dimensions of the field and the mass.
Knowledge of the renormalization functions also provides approximations for the critical exponents of phase transitions in certain universality classes of statistical models \cite{ZJ}.
So, the calculation of $\beta$, $\gamma$, and $\gamma_m$ to highest possible loop orders has impact beyond QFT.

The classical method to calculate these renormalization functions is based on momentum space. In many decades of research, very refined techniques were developed, most notably
the reduction to master integrals \cite{CTIBP,TIBP} after infrared $R^\ast$-reduction \cite{CSR,CTR}.
An overview over the classical methods is in \cite{KP6loopbeta} in the context of $\phi^4$ theory.
Outside $\phi^4$ theory, the beta function is typically known to five loops \cite{5lQCD1,5lQCD2,5lQCD3,5lphi3,5lQCD4,5lQCD5,5lQCD6,5lQCD7,classphi3,5lQCD8,5lQCD9,5lQCD10}.

A decade ago, the theory of graphical functions was developed by the author to calculate Feynman periods in four dimensions (residues of scalar Feynman integrals with
a single logarithmic divergence) \cite{gf}.
Graphical functions are single-valued functions on the complex plane with singularities at 0 and 1 that are defined by position space three-point integrals \cite{par}.
Many graphical functions can be expressed in the function space of generalized single-valued hyperlogarithms (GSVHs) \cite{GSVH}.

Using the theory of graphical functions, it was possible to calculate a significant number of Feynman periods.
With these results, the author discovered a connection between QFT and motivic Galois theory, a deep and rich mathematical structure which generalizes Galois theory to higher dimensions
in the context of algebraic integrals (see e.g.\ \cite{Bcoact1,Bcoact2}). The discovery led to the formulation of the coaction conjectures in \cite{coaction}.
Motivic Galois theory (also named `the coaction principle' or `the cosmic Galois group') is now a classic field in modern QFT (see e.g.\ \cite{Cosmic,motg2} and the references therein).

In collaboration with Michael Borinsky, the theory of graphical functions was generalized to even dimensions $\geq4$ \cite{gfe}.
It became possible to address $\phi^3$ theory which has a spin zero boson with a cubic self-interaction in six dimensions.
The graphical function method provided a complete list of Feynman periods up to six loops and partial results up to nine loops \cite{phi3,Shlog}.
All results are consistent with the coaction conjectures.

With graphical functions, dimensionally regularized $\phi^3$ theory could readily be renormalized up to five loops \cite{5lphi3}. The result was confirmed by a classical
calculation \cite{classphi3}. The quest for six loops started right after the five loop result was obtained. A successful calculation of the field anomalous dimension
(being significantly easier than the beta function), was finished in 2022 and published together with the seven loop beta function in $\phi^4$ theory \cite{7loops}.
Note that the result ($\zeta(5,3)=\sum_{k_1>k_2\geq1}\frac1{k_1^5k_2^3}$)
\begin{align}\label{6loopgamma}
\gamma_6^{\phi^3}&=-\frac{1567}{72}\zeta(9)-\zeta(3)^3-\frac{21}{10}\zeta(5,3)-\frac{2}{3}\zeta(5)\zeta(3)+\frac{209}{324000}\pi^8-\frac{25967}{2304}\zeta(7)
-\frac{13}{14400}\pi^4\zeta(3)\nonumber\\
&-\,\frac{11333}{10368}\zeta(3)^2+\frac{25637}{7838208}\pi^6+\frac{708913}{77760}\zeta(5)+\frac{1378253}{223948800}\pi^4+\frac{574643}{46656}\zeta(3)
+\frac{29506113557}{9674588160}\\
&=-0.331182447708\ldots\nonumber
\end{align}
has a somewhat unexpected negative sign.

The calculation of the six loop beta function needed one more year. The result
\begin{align}
\beta_6^{\phi^3}&=\frac{245045}{144}\zeta(9)+37\zeta(3)^3+\frac{3357}{40}\zeta(5,3)-\frac{11}{3}\zeta(5)\zeta(3)-\frac{81733}{2016000}\pi^8-\frac{456443}{1152}\zeta(7)
+\frac{99}{800}\pi^4\zeta(3)\nonumber\\
&-\,\frac{2425}{384}\zeta(3)^2+\frac{176425}{2612736}\pi^6-\frac{24878747}{34560}\zeta(5)+\frac{42654751}{74649600}\pi^4-\frac{85523425}{186624}\zeta(3)-\frac{173655397121}{3224862720}\\
&=\,-241.455497609497\ldots\nonumber
\end{align}
was proposed in a talk at RadCor 2023 as a preliminary result \cite{Radcortalk}.
Now, the result has undergone significant further testing (without corrections). Including $O(N)$ symmetry factors, it is contained in the Maple package
{\tt HyperlogProcedures} \cite{Shlog}. The arXiv version of this article is supplemented by a human readable file with the renormalization functions in $\phi^3$ theory and in $\phi^4$ theory.

The theory of graphical functions is explained in \cite{5lphi3,7loops}. The calculation of $\beta$ and $\gamma_m$ has significant overlap with the calculation of $\gamma$
from the six loop two-point function in \cite{7loops}.
To avoid copying material, we consider this article as an extension of \cite{7loops} rather then an independent paper.

We still try to keep the article self contained on a superficial level.
We focus on the $\phi^3$ three-point function and explain the calculation of the beta function from a certain truncated limit in the next Section.
In Section \ref{sect:phi3}, we explain how to calculate this truncated three-point function.
Finally, in Section \ref{sect:results}, we give the full results for the beta function, the field anomalous dimension and the mass anomalous dimension.

In this article, we do not calculate critical exponents for percolation or the Lee-Yang edge singularity \cite{5lphi3}.

First results for graphical functions in theories with particles of positive spin were developed in \cite{gft}.
In particular, it is shown that a dimensional shift mechanism maps calculations in four dimensions to even dimensions $\geq4$.
The author plans to extend applications of graphical functions to gauge theories with the aim to calculate six loop renormalization functions in QED and QCD.

\section*{Acknowlegements}
The author thanks Sven-Olaf Moch for support. During most of the work on this project the author was financed by the DFG grants SCHN~1240/2 and /3.

\section{Renormalization}\label{sect:ren}
We work in dimension
\begin{equation}
D=2\lambda+2=6-\epsilon
\end{equation}
and use a straight forward approach to renormalization. For each Feynman graph, we calculate the $\epsilon$-expansion of the Feynman integral to
the needed order and extract the $Z$ factors $Z_\phi,Z_g,Z_m$ which scale the bare values of the field $\phi$, the coupling $g$, and the mass $m$, see e.g.\ \cite{IZ}.
There exist much more refined techniques for calculating renormalization functions in momentum space (such as infrared rearrangement in the $R^\ast$ operation).
These techniques are to some extent builtin features of the graphical functions method.

For the beta function, we consider the three-point function and restrict ourselves to one particle irreducible (1PI) graphs.
We attach the three external legs to the external vertices $z_0$, $z_1$, $z_2$ and send one external vertex (say $z_2$) to infinity.
This is strictly speaking not necessary because at low loop orders we can calculate the three-point integrals with graphical functions.
However, the full structure of the three-point function is insignificant for renormalization. The limit $z_2\to\infty$ has the same $Z$ factors and is much easier to calculate.
Taking this limit is very efficient to reach high loop orders.

\begin{figure}
\begin{tikzpicture}
\begin{scope}[local bounding box=G]
    \filldraw[fill=black!10] (0,0) circle (1);
    \fill (-1,0) circle (2pt);
    \fill (0,1) circle (2pt);
    \fill (1,0) circle (2pt);
    \fill (-2,0) circle (2pt) node[anchor=north] {$z_0$};
    \fill (0,2) circle (2pt) node[anchor=west] {$z_2$};
    \fill (2,0) circle (2pt) node[anchor=north] {$z_1$};
    \draw (-1,0) -- (-2,0);
    \draw (0,1) -- (0,2);
    \draw (1,0) -- (2,0);
\end{scope}
    \node[below=0.2 of G] {$G$};
    \node at (2.9,0) {$\rightarrow$};

\begin{scope}[xshift=170,local bounding box=Gtrunc]
    \filldraw[fill=black!10] (0,0) circle (1);
    \fill (-1,0) circle (2pt);
    \fill (0,1) circle (2pt);
    \fill (1,0) circle (2pt);
    \fill (-2,0) circle (2pt) node[anchor=north] {$z_0$};
    \fill (2,0) circle (2pt) node[anchor=north] {$z_1$};
    \draw (-1,0) -- (-2,0);
    \draw (1,0) -- (2,0);
\end{scope}
    \node[below=0.2 of Gtrunc] {$G_{\mathrm{truncated}}$};
    \node at (8.9,0) {$\rightarrow$};

\begin{scope}[xshift=310,local bounding box=G3pt]
    \filldraw[fill=black!10] (0,0) circle (1);
    \fill (-1,0) circle (2pt);
    \fill (0,1) circle (2pt);
    \fill (1,0) circle (2pt);
    \draw (-1,0) .. controls (-1,2) and (1,2) .. (1,0);
\end{scope}
    \node[below=0.2 of G3pt] {$G_{\mathrm{3pt}}$};
    \node at (11.9,1.2) {$\nu_{\mathrm{3pt}}$};

\end{tikzpicture}
\caption{The reduction of the three-point graph $G$ to the period graph $G_{\mathrm{3pt}}$
with glued edge of weight $\nu_{\mathrm{3pt}}=h_G\epsilon/2\lambda$.}
\label{fig:3ptred}
\end{figure}

Because the three-point function is 1PI, the limit $z_2\to\infty$ has trivial scaling with a coefficient that corresponds to the truncated three-point function with deleted vertex $z_2$
\cite{7loops}. The deletion of $z_2$ converts the three-point function into an effective two-point
function. This two-point function is symmetric under $z_0\leftrightarrow z_1$ and has trivial scaling behavior. Its coefficient is a scalar integral (depending on the regulator $\epsilon$)
which is a Feynman period in the sense of \cite{numfunct}. This Feynman period can efficiently calculated with graphical functions using the freedom to choose new vertices $z_0$, $z_1$, $z_2$
in the graph.
We obtain a reduction to a period graph $G_{\mathrm{3pt}}$ which is $G\backslash z_2$ with an extra edge closing the half-edges attached to $z_0$ and $z_1$. The extra edge has weight
\begin{equation}\label{eq:wt3pt}
\nu_{\mathrm{3pt}}=\frac{h_G\epsilon}{2\lambda}
\end{equation}
while all other edges have weight $1$; see Figure \ref{fig:3ptred}. Here, $h_G$ is the number of loops (independent cycles) in $G$. We obtain
\begin{equation}\label{eq:3ptred}
\lim_{z_2\to\infty}||z_2||^{2\lambda}A_G(z_0,z_1,z_2)=A_{G\backslash z_2}(z_0,z_1)=
\frac{P_{G_{\mathrm{3pt}}}||z_1-z_0||^{-2+(h_G+1)\epsilon}}{\Gamma(\lambda)^2\lambda\nu_{\mathrm{3pt}}(\lambda\nu_{\mathrm{3pt}}+1)
(\lambda-\lambda\nu_{\mathrm{3pt}})(\lambda-1-\lambda\nu_{\mathrm{3pt}})}.
\end{equation}
The denominator in (\ref{eq:3ptred}) comes from integrating out the external edges attached to $z_0$ and $z_1$ in position space using
\begin{equation}\label{eq:nu1red}
\begin{tikzpicture}
    \fill (-1,0) circle (2pt);
    \fill (0,0) circle (2pt);
    \fill (1,0) circle (2pt);
    \draw (-1,0) -- (1,0);
    \node at (-0.5,0.2) {$\scriptstyle\nu$};
    \node at (0.5,0.2) {$\scriptstyle1$};
    \node at (3.7,0) {$\displaystyle=\;\frac1{\Gamma(\lambda)(\lambda(1-\nu)+1)(\lambda\nu-1)}$};
    \fill (6.3,0) circle (2pt);
    \fill (7.3,0) circle (2pt);
    \draw (6.3,0) -- (7.3,0);
    \node at (6.8,0.3) {$\scriptstyle\nu-\frac1\lambda$};
    \node at (7.6,-0.1) {,};
\end{tikzpicture}
\end{equation}
where scaling weights of propagators are indicated above the edges. For the two-point function we obtain
\begin{equation}\label{eq:2ptred}
A_G(z_0,z_1)=\frac{P_{G_{\mathrm{2pt}}}||z_1-z_0||^{-4+(h_G+1)\epsilon}}{\Gamma(\lambda)^2\lambda\nu_{\mathrm{3pt}}(\lambda\nu_{\mathrm{3pt}}-1)
(\lambda-\lambda\nu_{\mathrm{3pt}})(\lambda+1-\lambda\nu_{\mathrm{3pt}})},
\end{equation}
where $G_{\mathrm{2pt}}$ is the corresponding two-point period which has a glued edge of weight
\begin{equation}\label{eq:wt2pt}
\nu_{\mathrm{2pt}}=\frac{-2+h_G\epsilon}{2\lambda}.
\end{equation}
For the $\ell$ loop result, we need to calculate 1PI graphs of loop orders $k\leq\ell$ to order $\epsilon^{\ell-k-1}$. We determine the Laurent expansions of $Z_\phi,Z_g,Z_m$ such
that the result is regular for $\epsilon\to0$.

Because we work with leg-fixed Feynman graphs, we get $3!/2=3$ times the number of Feynman graph of a non-leg-fixed approach.
By graph isomorphism the exact number is a bit smaller. At loop order six in $\phi^3$ theory, one obtains $37307$ non-isomorphic graphs of Feynman periods which have to be
evaluated to order $\epsilon^{-1}$. In practice, the proliferation of graphs is not problematic because in most cases the calculations for permuted external vertices
are either identical or very similar. Graphical functions are quite efficient, so that computation time is not an issue. The difficulty of the calculation is to find
ways to calculate the few hard topologies; see Section \ref{sect:phi3}.

To extract critical exponents for both the Lee–Yang and percolation problems, it is necessary to include combinatorial symmetry factors.
The calculation of the symmetry factors is explained in \cite{5lphi3}.

From the calculation of the anomalous field dimension $\gamma$, we know $Z_\phi$ to loop order six \cite{7loops}.
The $Z$-factors $Z_g$ and $Z_m$ are determined such that the truncated three-point function has no poles in $\epsilon$ (with different symmetry factors).

The results for the renormalization functions are available in {\tt HyperlogProcedures} \cite{Shlog} which also includes $\epsilon$-expansions for all individual graphs.
As consistency test and to prepare future higher loop orders, in many cases the $\epsilon$-expansion is one order higher than
what is necessary for the six loop results. Currently, however, the calculation of the seven loop renormalization functions in $\phi^3$ theory is out of reach.

\section{The truncated three-point function}\label{sect:phi3}
\begin{figure}
\centering
\includegraphics{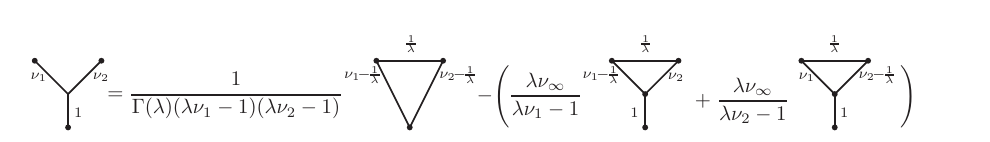}
\caption{The Y$\nabla\overline{\mathrm{Y}}\,\overline{\mathrm{Y}}$-identity, where $\nu_\infty=1+\frac2\lambda-\nu_1-\nu_2$.
This identity is particularly useful for $\nu_1=\nu_2=1$ in $6-\epsilon$ dimensions.}
\label{fig:YDYY}
\end{figure}

The main tools for calculating the period $P_{G_{\mathrm{3pt}}}$ are completion and local integration by parts (IBP) as explained in Section 8 of \cite{7loops}.
In particular, the Y$\nabla\overline{\mathrm{Y}}\,\overline{\mathrm{Y}}$-identity depicted in Figure \ref{fig:YDYY} is of central importance.
It is an extension of the uniqueness identity (the star-triangle identity) to non-integer dimensions \cite{unique1,DY,K2,unique2}. Note that the two graphs with $\overline{\mathrm{Y}}$ topology
on the right hand side are suppressed by one power in $\epsilon$ if $\nu_1+\nu_2=2+O(\epsilon)$.

Integration by parts is only used locally at fixed vertices. It is not possible to apply the Laporta-algorithm \cite{La,LaRe}
to graphs of loop order six (also see \cite{IBP} and the references therein).
A more systematic use of IBP equations was employed for the calculation of the residues of (primitive) three-point graphs without subdivergences \cite{phi3}.

Truncation of the edge attached to $z_2$ gives a vertex of degree two in the graph $G_{\mathrm{3pt}}$.
This vertex can be eliminated with (\ref{eq:nu1red}),

\begin{equation}\label{eq:11red}
\begin{tikzpicture}
    \fill (-1,0) circle (2pt);
    \fill (0,0) circle (2pt);
    \fill (1,0) circle (2pt);
    \draw (-1,0) -- (1,0);
    \node at (-0.5,0.2) {$\scriptstyle1$};
    \node at (0.5,0.2) {$\scriptstyle1$};
    \node at (2.6,0) {$\displaystyle=\;\frac1{\Gamma(\lambda)(\lambda-1)}$};
    \fill (4,0) circle (2pt);
    \fill (5,0) circle (2pt);
    \draw (4,0) -- (5,0);
    \node at (4.5,0.3) {$\scriptstyle1-\frac1\lambda$};
    \node at (5.3,-0.1) {.};
\end{tikzpicture}
\end{equation}

\begin{figure}
\begin{tikzpicture}
\begin{scope}[local bounding box=G1]
    \filldraw[fill=black!10] (0,0) circle (0.5);
    \node at (0,0) {$G'_{\mathrm{3pt}}$};
    \fill (-0.5,0) circle (2pt);
    \fill (0.5,0) circle (2pt);
    \fill (-1,0) circle (2pt);
    \fill (1,0) circle (2pt);
    \fill (0,0.75) circle (0pt) node[anchor=south] {$\nu_{\mathrm{3pt}}$};
    \draw (-1,0) .. controls (-1,1) and (1,1) .. (1,0);
    \draw (-1,0) .. controls (-1,-1) and (1,-1) .. (1,0);
    \draw (-1,0) -- (-0.5,0);
    \draw (1,0) -- (0.5,0);
\end{scope}
    \node[below=0 of G1] {Type $1$};

\begin{scope}[xshift=100,local bounding box=G2]
    \filldraw[fill=black!10] (0,0) circle (0.5);
    \node at (0,0) {$G'_{\mathrm{2pt}}$};
    \fill (-0.5,0) circle (2pt);
    \fill (0.5,0) circle (2pt);
    \fill (-1,0) circle (2pt);
    \fill (1,0) circle (2pt);
    \fill (0,0.75) circle (0pt) node[anchor=south] {$\nu_{\mathrm{3pt}}$};
    \fill (0,-0.75) circle (2pt);
    \draw (-1,0) .. controls (-1,1) and (1,1) .. (1,0);
    \draw (-1,0) .. controls (-1,-1) and (1,-1) .. (1,0);
    \draw (-1,0) -- (-0.5,0);
    \draw (1,0) -- (0.5,0);
\end{scope}
    \node[below=0 of G2] {Type $2$};

\begin{scope}[xshift=200,local bounding box=G3]
    \filldraw[fill=black!10] (0,0) circle (0.5);
    \fill (-0.5,0) circle (2pt);
    \fill (0.47,0.2) circle (2pt);
    \fill (0.47,-0.2) circle (2pt);
    \fill (0,-0.5) circle (2pt) node[anchor=north west] {$2$};
    \fill (-1,0) circle (2pt) node[anchor=east] {$0$};
    \fill (1,0) circle (2pt) node[anchor=west] {$1$};
    \fill (0,0.75) circle (0pt) node[anchor=south] {$\nu_{\mathrm{3pt}}$};
    \fill (-0.6,-0.55) circle (2pt);
    \draw (-1,0) .. controls (-1,1) and (1,1) .. (1,0);
    \draw (-1,0) .. controls (-1,-0.5) and (0,-1) .. (0,-0.5);
    \draw (-1,0) -- (-0.5,0);
    \draw (1,0) -- (0.47,-0.2);
    \draw (1,0) -- (0.47,0.2);
\end{scope}
    \node[below=0 of G3] {Type $3$};

\begin{scope}[xshift=300,local bounding box=G4]
    \filldraw[fill=black!10] (0,0) circle (0.5);
    \fill (0.47,0.2) circle (2pt);
    \fill (0.47,-0.2) circle (2pt);
    \fill (-0.47,0.2) circle (2pt);
    \fill (-0.47,-0.2) circle (2pt);
    \fill (0.3,-0.4) circle (2pt) node[anchor=north west] {$3$};
    \fill (-0.3,-0.4) circle (2pt) node[anchor=north east] {$2$};
    \fill (0,-0.85) circle (2pt);
    \fill (-1,0) circle (2pt) node[anchor=east] {$0$};
    \fill (1,0) circle (2pt) node[anchor=west] {$1$};
    \fill (0,0.75) circle (0pt) node[anchor=south] {$\nu_{\mathrm{3pt}}$};
    \draw (-1,0) .. controls (-1,1) and (1,1) .. (1,0);
    \draw (-0.3,-0.4) .. controls (-0.2,-1) and (0.2,-1) .. (0.3,-0.4);
    \draw (-1,0) -- (-0.47,-0.2);
    \draw (-1,0) -- (-0.47,0.2);
    \draw (1,0) -- (0.47,-0.2);
    \draw (1,0) -- (0.47,0.2);
\end{scope}
    \node[below=0 of G4] {Type $4$};
    \end{tikzpicture}
\caption{The four types of the period graph $G_{\mathrm{3pt}}$.
In types 3 and 4, the two-chain connects to one or none of the vertices of the glued edge
(and they are not of Type 1).}
\label{fig:types}
\end{figure}

We classify the three-point graphs $G_{\mathrm{3pt}}$ into four types according to Figure \ref{fig:types}.

Types $1$ and $2$ are the cases when an edge or the two-chain at the truncated vertex $z_2$ are attached to the vertices of the glued edge.
In these cases, one gets a reduction to lower loop order and higher degree in $\epsilon$. The glued edge combines with the edge(s) in the lower arcs in Figure \ref{fig:types} and
we obtain from (\ref{eq:nu1red}),
\begin{align}
P_{G_{\mathrm{3pt}}}^{\mathrm{Type\,1}}&=\frac{P_{G'_{\mathrm{3pt}}}}{\Gamma(\lambda)^2(1-\lambda\nu_{\mathrm{3pt}})(2-\lambda\nu_{\mathrm{3pt}})
(\lambda-1+\lambda\nu_{\mathrm{3pt}})(\lambda-2+\lambda\nu_{\mathrm{3pt}})}\nonumber\\
P_{G_{\mathrm{3pt}}}^{\mathrm{Type\,2}}&=\frac{P_{G'_{\mathrm{2pt}}}}{\Gamma(\lambda)^3(\lambda-1)(2-\lambda\nu_{\mathrm{3pt}})(3-\lambda\nu_{\mathrm{3pt}})
(\lambda-2+\lambda\nu_{\mathrm{3pt}})(\lambda-3+\lambda\nu_{\mathrm{3pt}})}.
\end{align}
On the right hand sides, the graphs $G'_{\mathrm{3pt}}$ and $G'_{\mathrm{2pt}}$ are the internal reduced three-point and two-point period graphs.
These graphs are shaded in Figure \ref{fig:types} with an additional glued edge of weight (\ref{eq:wt3pt}) or (\ref{eq:wt2pt}), respectively, with loop order $h_G-1$.
Because $\lambda-2+\lambda\nu_{\mathrm{3pt}}=(h_G-1)\epsilon/2$, we need to calculate the reduced graphs to one extra order in $\epsilon$.
The result can be taken from the contribution of loop order $h_G-1$ to the truncated three-point function or the two-point function because this contribution has to be calculated to
one order in $\epsilon$ higher than the contribution to loop order $h_G$; see Section \ref{sect:ren}.

Types $3$ and $4$ are the cases when no edges exist that directly connect the vertices of the glued edge.
Then, we distinguish between the case that the truncated two-chain is attached to a vertex of the glued edge (Type $3$) or not (Type $4$).
Because the glued edge has a weight that vanishes for $\epsilon=0$, one can efficiently use the local IBP identity at its vertices,
\begin{equation}\label{eq:IBP}
\begin{tikzpicture}
    \fill (-0.5,0) circle (2pt);
    \fill (0.5,0) circle (2pt);
    \fill (0.5,1) circle (2pt);
    \fill (0.5,-1) circle (2pt);
    \draw (-0.5,0) -- (0.5,0) -- (0.5,1);
    \draw (0.5,0) -- (0.5,-1);
    \node at (0,0.2) {$\scriptstyle1$};
    \node at (0.2,+0.7) {$\scriptstyle\nu_{\mathrm{3pt}}$};
    \node at (0.3,-0.7) {$\scriptstyle\nu_1$};
    \node at (3.2,0) {$=\;\displaystyle\frac1{\Gamma(\lambda)(\lambda\nu_1-1)(\lambda\nu_\infty-1)}$};
    \fill (5.5,0) circle (2pt);
    \fill (6.5,1) circle (2pt);
    \fill (6.5,-1) circle (2pt);
    \draw (6.5,1) -- (5.5,0) -- (6.5,-1);
    \node at (5.9,0.7) {$\scriptstyle\nu_{\mathrm{3pt}}$};
    \node at (5.75,-0.7) {$\scriptstyle\nu_1-\frac1\lambda$};
    \node at (7.9,0) {$-\;\displaystyle\frac{\lambda\nu_{\mathrm{3pt}}}{\lambda\nu_1-1}$};
    \fill (9,0) circle (2pt);
    \fill (10,0) circle (2pt);
    \fill (10,1) circle (2pt);
    \fill (10,-1) circle (2pt);
    \draw (9,0) -- (10,0) -- (10,1);
    \draw (10,0) -- (10,-1);
    \node at (9.5,0.2) {$\scriptstyle1$};
    \node at (9.5,0.7) {$\scriptstyle\nu_{\mathrm{3pt}}+\frac1\lambda$};
    \node at (9.6,-0.7) {$\scriptstyle\nu_1-\frac1\lambda$};
    \node at (11.5,0) {$-\;\displaystyle\frac{\lambda\nu_{\mathrm{3pt}}}{\lambda\nu_\infty-1}$};
    \fill (12.7,0) circle (2pt);
    \fill (13.7,0) circle (2pt);
    \fill (13.7,1) circle (2pt);
    \fill (13.7,-1) circle (2pt);
    \draw (12.7,0) -- (13.7,0) -- (13.7,1);
    \draw (13.7,0) -- (13.7,-1);
    \draw (13.7,-1) .. controls (14.7,-1) and (14.7,1) .. (13.7,1);
    \node at (13.2,0.2) {$\scriptstyle1$};
    \node at (13.2,0.7) {$\scriptstyle\nu_{\mathrm{3pt}}+\frac1\lambda$};
    \node at (13.5,-0.7) {$\scriptstyle\nu_1$};
    \node at (14.2,0) {$\scriptstyle-\frac1\lambda$};
\end{tikzpicture}
\end{equation}
where $\nu_\infty=1+\frac2\lambda-\nu_1-\nu_{\mathrm{3pt}}$.

In both types, 3 and 4, there exists a vertex of the glued edge with $\nu_1=1$. In this setup, (\ref{eq:IBP}) is particularly useful because the two rightmost graphs are suppressed
by one order in $\epsilon$. We obtain

\begin{equation}\label{eq:IBP1}
\begin{tikzpicture}
    \filldraw[fill=black!10] (0,0) circle (0.5);
    \fill (-0.47,0.2) circle (2pt);
    \fill (-0.47,-0.2) circle (2pt);
    \fill (-1,0) circle (2pt);
    \fill (0.5,0) circle (2pt);
    \draw (-1,0) .. controls (-1,1) and (0.5,1) .. (0.5,0);
    \draw (-1,0) -- (-0.47,-0.2);
    \draw (-1,0) -- (-0.47,0.2);
    \node at (-0.25,0.95) {$\scriptstyle\nu_{\mathrm{3pt}}$};
    \node at (-0.75,0.28) {$\scriptstyle1$};
    \node at (-0.75,-0.28) {$\scriptstyle1$};
    \node at (2.8,0) {$=\;\displaystyle\frac1{\Gamma(\lambda)(\lambda-1)(1-\lambda\nu_{\mathrm{3pt}})}$};
    \filldraw[fill=black!10] (6,0) circle (0.5);
    \fill (5.53,0.2) circle (2pt);
    \fill (5.53,-0.2) circle (2pt);
    \fill (6.5,0) circle (2pt);
    \draw (5.53,-0.2) .. controls (5,-0.2) and (5,0.2) .. (5.53,0.2);
    \draw (5.53,0.2) .. controls (5.53,1) and (6.55,1) .. (6.5,0);
    \node at (6,0.95) {$\scriptstyle\nu_{\mathrm{3pt}}$};
    \node at (5.2,0.4) {$\scriptstyle1-\frac1\lambda$};
    \node at (7.5,0) {$-\;\displaystyle\frac{\lambda\nu_{\mathrm{3pt}}}{\lambda-1}$};
    \filldraw[fill=black!10] (9.5,0) circle (0.5);
    \fill (9.03,0.2) circle (2pt);
    \fill (9.03,-0.2) circle (2pt);
    \fill (8.5,0) circle (2pt);
    \fill (10,0) circle (2pt);
    \draw (8.5,0) .. controls (8.5,1) and (10,1) .. (10,0);
    \draw (8.5,0) -- (9.03,-0.2);
    \draw (8.5,0) -- (9.03,0.2);
    \node at (9.25,0.95) {$\scriptstyle\nu_{\mathrm{3pt}}+\frac1\lambda$};
    \node at (8.75,0.28) {$\scriptstyle1$};
    \node at (8.7,-0.35) {$\scriptstyle1-\frac1\lambda$};
    \node at (11.3,0) {$-\;\displaystyle\frac{\lambda\nu_{\mathrm{3pt}}}{1-\lambda\nu_{\mathrm{3pt}}}$};
    \filldraw[fill=black!10] (13.5,0) circle (0.5);
    \fill (13.03,0.2) circle (2pt);
    \fill (13.03,-0.2) circle (2pt);
    \fill (12.5,0) circle (2pt);
    \fill (14,0) circle (2pt);
    \draw (12.5,0) .. controls (12.5,1) and (14,1) .. (14,0);
    \draw (13.03,-0.2) .. controls (13.03,-1) and (14,-1) .. (14,0);
    \draw (12.5,0) -- (13.03,-0.2);
    \draw (12.5,0) -- (13.03,0.2);
    \node at (13.25,0.95) {$\scriptstyle\nu_{\mathrm{3pt}}+\frac1\lambda$};
    \node at (12.75,0.28) {$\scriptstyle1$};
    \node at (12.75,-0.28) {$\scriptstyle1$};
    \node at (13.9,-0.8) {$\scriptstyle-\frac1\lambda$};   
\end{tikzpicture}
\end{equation}

In Type 3, we have an identity with $\nu_1=1-1/\lambda$. In this case, the vertex is singular and produces terms with $\epsilon^{-1}$. Also here,
(\ref{eq:IBP}) can be useful although a simplification is not evident.

Another option is to use completion and thereafter local IBP at vertex $\infty$ as explained in Sections 8.1 and 8.2 of \cite{7loops}.
Completion adds a vertex $\infty$ which connects to all other vertices such that the weighted degree of every vertex is $D/\lambda$. Local IBP gives identities for sums of graphs which
differ by locally adding and subtracting triangles with edges of weights $1/\lambda$, $-1/\lambda$, $1/\lambda$. Local IBP at $\infty$ is formulated in terms of graphs
$G^{ij}_{kl}$ which are $G_{\mathrm{3pt}}$ with an edge $ij$ of weight $1/\lambda$ and an edge $kl$ of weight $-1/\lambda$.
In our setup, $ij=01$ are the vertices of the glued edge. This edge has weight $\nu_{\mathrm{3pt}}+1/\lambda$ in $G^{01}_{kl}$ for $k\ell\neq01$ while $G^{01}_{01}=G_{\mathrm{3pt}}$.
In graphs of types 3 and 4, we label the vertices where the two-chain connects to the rest of the graph with $2$ and $3$ (see Figure \ref{fig:types}).
We define $P_{G^{01}_{kk}}=0$ for all $k$ and use the abbreviations
\begin{equation}
\alpha=2+\epsilon,\quad\beta=2-h_G\epsilon,\quad\gamma=2-(h_G-3)\epsilon.
\end{equation}
For graphs of Type 3 we get the identities
\begin{align}
P_{G_{\mathrm{3pt}}}^{\mathrm{Type\,3}}&=P_{G^{01}_{1i}}-\frac\alpha\beta(P_{G^{01}_{02}}-P_{G^{01}_{2i}})+\frac{\beta+2}\beta P_{G^{01}_{0i}}
-\frac\epsilon\beta\sum_{k\geq3}(P_{G^{01}_{0k}}-P_{G^{01}_{ik}})\nonumber\\
&=P_{G^{01}_{0i}}-\frac\alpha{\beta+2}(P_{G^{01}_{12}}-P_{G^{01}_{2i}})+\frac\beta{\beta+2} P_{G^{01}_{1i}}-\frac\epsilon{\beta+2}\sum_{k\geq3}(P_{G^{01}_{1k}}-P_{G^{01}_{ik}}),
\end{align}
with $i\geq2$. For general $i$, the right hand sides are not simpler than $P_{G_{\mathrm{3pt}}}$. However, it may happen that there exists a particularly convenient value for $i$ such
that the right hand side can be calculated in a case where $P_{G_{\mathrm{3pt}}}$ is not accessible by other methods.

For graphs of Type 4 we get two duality formulae for $i=1-j=0,1$,
\begin{equation}
P_{G_{\mathrm{3pt}}}^{\mathrm{Type\,4}}=\frac{\alpha^2}{\beta\gamma}P_{G^{01}_{23}}-\frac{3\alpha\epsilon}{\beta\gamma}(P_{G^{01}_{i2}}+P_{G^{01}_{i3}})-\epsilon\sum_{k\geq4}
\Big(\frac1\beta P_{G^{01}_{ik}}+\frac1\gamma P_{G^{01}_{jk}}\Big)-\frac{\epsilon^2}{\beta\gamma}\sum_{4\leq k<\ell}P_{G^{01}_{kl}}
\end{equation}
in which, up to terms of order $\epsilon$, the weights of the edges $01$ and $23$ are swapped.

There exists a plethora of similar formulae that can be useful for various topologies. Some of these formulae were implemented in {\tt HyperlogProcedures} \cite{Shlog}
to complete the calculation of the six loop beta function in $\phi^3$ theory.

\section{Results}\label{sect:results}

In this section we summarize the six loop results for the renormalization functions of $\phi^3$ theory in the minimal subtraction scheme.
The results were calculated using {\tt HyperlogProcedures} \cite{Shlog} over several weeks on a server with 36 cores and 780 GB RAM.
We confirm the five loop results for the Lee-Yang edge singularity in \cite{5lphi3} and obtain for the beta function $\beta^{\phi^3}$,
the anomalous field dimension $\gamma^{\phi^3}$, and the mass anomalous dimension $\gamma_m^{\phi^3}$,

\begin{align}\label{full6loopbeta}
\beta^{\phi^3}\!(g)&=\Big(\frac{245045}{144}\zeta(9)+37\zeta(3)^3+\frac{3357}{40}\zeta(5,3)-\frac{11}{3}\zeta(5)\zeta(3)-\frac{81733}{2016000}\pi^8-\frac{456443}{1152}\zeta(7)
+\frac{99}{800}\pi^4\zeta(3)\nonumber\\
&-\,\frac{2425}{384}\zeta(3)^2+\frac{176425}{2612736}\pi^6-\frac{24878747}{34560}\zeta(5)+\frac{42654751}{74649600}\pi^4-\frac{85523425}{186624}\zeta(3)
-\frac{173655397121}{3224862720}\Big)g^{13}\nonumber\\
&+\,\Big(\!-\frac{5495}{64}\zeta(7)+\frac{99}{16}\zeta(3)^2+\frac{5}{2016}\pi^6+\frac{151795}{3456}\zeta(5)-\frac{46519}{829440}\pi^4+\frac{366647}{6912}\zeta(3)
+\frac{102052031}{6718464}\Big)g^{11}\nonumber\\
&+\,\Big(\frac{5}{3}\zeta(5)+\frac{1}{192}\pi^4-\frac{4891}{864}\zeta(3)-\frac{3404365}{746496}\Big)g^9+\Big(\frac{5}{8}\zeta(3)+\frac{33085}{20736}\Big)g^7-\frac{125}{144}g^5
+\frac{3}{4}g^3\nonumber\\
&\approx-241.45550g^{13}+43.78256g^{11}-9.12961g^9+2.34682g^7-0.86806g^5+0.75g^3,
\end{align}

\begin{align}\label{full6loopgamma}
\gamma^{\phi^3}\!(g)&=\Big(\!-\frac{1567}{72}\zeta(9)-\zeta(3)^3-\frac{21}{10}\zeta(5,3)-\frac{2}{3}\zeta(5)\zeta(3)+\frac{209}{324000}\pi^8-\frac{25967}{2304}\zeta(7)
-\frac{13}{14400}\pi^4\zeta(3)\nonumber\\
&-\,\frac{11333}{10368}\zeta(3)^2+\frac{25637}{7838208}\pi^6+\frac{708913}{77760}\zeta(5)+\frac{1378253}{223948800}\pi^4+\frac{574643}{46656}\zeta(3)
+\frac{29506113557}{9674588160}\Big)g^{12}\nonumber\\
&+\,\Big(\frac{147}{64}\zeta(7)-\frac{25}{144}\zeta(3)^2-\frac{25}{54432}\pi^6+\frac{4471}{10368}\zeta(5)-\frac{5651}{2488320}\pi^4-\frac{56693}{62208}\zeta(3)
-\frac{16492987}{20155392}\Big)g^{10}\nonumber\\
&+\,\Big(\!-\frac{5}{18}\zeta(5)+\frac{7}{8640}\pi^4+\frac{35}{864}\zeta(3)+\frac{53449}{248832}\Big)g^8+\Big(\frac{1}{24}\zeta(3)-\frac{5195}{62208}\Big)g^6
+\frac{13}{432}g^4-\frac{1}{12}g^2\nonumber\\
&\approx-0.33118g^{12}-0.06420g^{10}+0.05438g^8-0.03342g^6+0.03009g^4-0.08333g^2,
\end{align}

\begin{align}\label{full6loopgammam}
\gamma_m^{\phi^3}\!(g)&=\Big(\!-\frac{248179}{144}\zeta(9)-38\zeta(3)^3\hspace{-0.7pt}-\frac{3441}{40}\zeta(5,3)+3\zeta(5)\zeta(3)+\frac{747301}{18144000}\pi^8\hspace{-0.7pt}
+\frac{886919}{2304}\zeta(7)-\frac{359}{2880}\pi^4\zeta(3)\nonumber\\
&+\,\frac{27071}{5184}\zeta(3)^2-\frac{251819}{3919104}\pi^6+\frac{45348875}{62208}\zeta(5)-\frac{316465}{559872}\pi^4+\frac{29273999}{62208}\zeta(3)
+\frac{13761807623}{241864704}\Big)g^{12}\nonumber\\
&+\,\Big(\frac{2821}{32}\zeta(7)-\frac{229}{36}\zeta(3)^2-\frac{5}{1701}\pi^6-\frac{225457}{5184}\zeta(5)+\frac{66953}{1244160}\pi^4-\frac{839129}{15552}\zeta(3)
-\frac{40331135}{2519424}\Big)g^{10}\nonumber\\
&+\,\Big(\!-\frac{35}{18}\zeta(5)-\frac{19}{4320}\pi^4+\frac{821}{144}\zeta(3)+\frac{445589}{93312}\Big)g^8+\Big(\!-\frac{7}{12}\zeta(3)-\frac{52225}{31104}\Big)g^6+\frac{97}{108}g^4
-\frac{5}{6}g^2\nonumber\\
&\approx 241.12432g^{12}-43.84676g^{10}+9.18399g^8-2.38024g^6+0.89815g^4-0.83333g^2.
\end{align}
The six loop results can be expressed in terms of transcendentals in $\phi^4$ theory \cite{BK,Census}.

The author did not attempt to obtain improved values for critical exponents for percolation or the Lee–Yang edge singularity.

\bibliographystyle{plain}
\renewcommand\refname{References}

\end{document}